\documentclass[twocolumn,citeautoscript,pra,superscriptaddress,amsmath,amssymb,8pt]{revtex4-1}
\usepackage{graphicx}
\usepackage{gensymb}
\usepackage{color}
\usepackage[dvipsnames]{xcolor}
\usepackage{float}
\usepackage{upgreek}
\usepackage{amsmath,amssymb}

\newcommand{\ket}[1]{\ensuremath{\left| #1 \right\rangle}}

\graphicspath{{C:/Users/Alex/Desktop/plots/hbn_irrad/}}

\begin{document}

\title{Optimisation of electron irradiation for creating spin ensembles in hexagonal boron nitride}

\author{Alexander J. Healey}
\email{alexander.healey2@rmit.edu.au}	
\affiliation{School of Science, RMIT University, Melbourne, VIC 3001, Australia}

\author{Priya Singh}
\affiliation{School of Science, RMIT University, Melbourne, VIC 3001, Australia}

\author{Islay O. Robertson}
\affiliation{School of Science, RMIT University, Melbourne, VIC 3001, Australia}

\author{Christopher Gavin}
\affiliation{School of Science, RMIT University, Melbourne, VIC 3001, Australia}

\author{Sam C. Scholten}
\affiliation{School of Physics, University of Melbourne, VIC 3010, Australia}
\affiliation{Centre for Quantum Computation and Communication Technology, School of Physics, University of Melbourne, VIC 3010, Australia}

\author{David A. Broadway}
\affiliation{School of Science, RMIT University, Melbourne, VIC 3001, Australia}

\author{Philipp Reineck}
\affiliation{School of Science, RMIT University, Melbourne, VIC 3001, Australia}

\author{Hiroshi Abe}
\affiliation{National Institutes for Quantum Science and Technology (QST), 1233 Watanuki, Takasaki, Gunma 370-1292, Japan}

\author{Takeshi Ohshima}
\affiliation{National Institutes for Quantum Science and Technology (QST), 1233 Watanuki, Takasaki, Gunma 370-1292, Japan}
\affiliation{Department of Materials Science, Tohoku University, 6-6-02 Aramaki-Aza, Aoba-ku, Sendai 980-8579, Japan}

\author{Mehran Kianinia}
\affiliation{School of Mathematical and Physical Sciences, University of Technology Sydney, Ultimo, NSW 2007, Australia}
\affiliation{ARC Centre of Excellence for Transformative Meta-Optical Systems, University of Technology Sydney, Ultimo, NSW 2007, Australia}

\author{Igor Aharonovich}
\affiliation{School of Mathematical and Physical Sciences, University of Technology Sydney, Ultimo, NSW 2007, Australia}
\affiliation{ARC Centre of Excellence for Transformative Meta-Optical Systems, University of Technology Sydney, Ultimo, NSW 2007, Australia}

\author{Jean-Philippe Tetienne}
\email{Jean-philippe.tetienne@rmit.edu.au}
\affiliation{School of Science, RMIT University, Melbourne, VIC 3001, Australia}

\begin{abstract}
Boron vacancy centre ($V_{\rm B}^-$) ensembles in hexagonal boron nitride (hBN) have attracted recent interest for their potential as two-dimensional solid-state quantum sensors. Irradiation is necessary for $V_{\rm B}^-$ creation, however, to date only limited attention has been given to optimising the defect production process, especially in the case of bulk irradiation with high-energy particles, which offers scalability through the potential for creating ensembles in large volumes of material. Here we systematically investigate the effect of electron irradiation by varying the dose delivered to a range of hBN samples, which differ in their purity, and search for an optimum in measurement sensitivity. We find that moderate electron irradiation doses ($\approx 5\times 10^{18}$~cm$^{-2}$) appear to offer the best sensitivity, and also observe a dependence on the initial crystal purity. These results pave the way for the scalable and cost-effective production of hBN quantum sensors, and provide insight into the mechanisms limiting $V_{\rm B}^-$ spin properties. 
\end{abstract}

\maketitle 
\section{Introduction}
The advent of optically active spin defects in hBN, in particular the negatively-charged boron vacancy centre ($V_{\rm B}^-$)~\cite{Gottscholl2020,Gottscholl2021,Gottscholl2021a}, has sparked excitement as van der Waals quantum sensors have opened up the possibility of moving towards atomic proximity with a target and for seamless integration with devices based on van der Waals heterostructures~\cite{Aharonovich2022,Vaidya2023}. Implementations based on single crystal flakes exoliated from high-quality bulk crystals~\cite{Healey2022a,Huang2022,Kumar2022} and commercially-available nanopowders~\cite{Robertson2023,Scholten2023} have been demonstrated, with proposed applications in fields ranging from condensed matter physics to biology~\cite{Vaidya2023}. Essential to the furtherance of hBN-based quantum sensing is arriving at reliable and scalable methods for creating high quality spin ensembles.

Low-energy ($<100~$keV) ion implantation has been a popular method for creating $V_{\rm B}^-$ ensembles so far, partly because it creates a predictable and constrained sensing layer~\cite{Kianinia2020}. However, this depth confinement is also a weakness since a single implantation run only activates an effectively 2D layer. Conversely, electron irradiation, at MeV energies, creates vacancies evenly throughout an irradiated volume that can extend to the millimetre scale~\cite{Campbell2002}. Electron irradiation, therefore, is naturally well-suited to addressing the scalability problem by allowing the creation of $V_{\rm B}^-$ ensembles throughout entire bulk crystals or a large volume of nanoflakes. In principle, a localised (2D) layer can also still be realised by exfoliating a sufficiently thin flake from a bulk crystal~\cite{Durand2023}. Additionally, electron irradiation is accessible as a commercial service, in contrast to neutron irradiation, which has also been employed for this purpose~\cite{Gottscholl2020,Toledo2018}. 

In this work, we irradiate hBN samples of varying purity with 2~MeV electrons over a range of doses between $2~\times 10^{16}$ and $2~\times 10^{19}$~cm$^{-2}$. This study design allows us to systematically examine the effect of lattice damage (proportional to irradiation dose) as well as initial crystal purity on $V_{\rm B}^-$ spin properties and measurement sensitivity.

Previous works considering $V_{\rm B}^-$ creation have focussed primarily on high-density ion implantation~\cite{Kianinia2020} and neutron irradiation~\cite{Gottscholl2020}, and have not explicitly investigated measurement sensitivity, which is an experiment-dependent quantity that generally scales in nontrivial ways with defect density in similar solid-state systems~\cite{Healey2020}. Furthermore, the dependence on hBN purity has also not been investigated so far. Given the dominance of the dense nuclear spin bath present in hBN in setting the spin coherence of $V_{\rm B}^-$ defects~\cite{Haykal2022}, it is not \textit{a priori} clear whether defects related to irradiation damage or crystal impurities should have a strong effect on ensemble spin properties and hence measurement sensitivity.
\begin{figure*}[hbt!]
\centering
\includegraphics[scale=1]{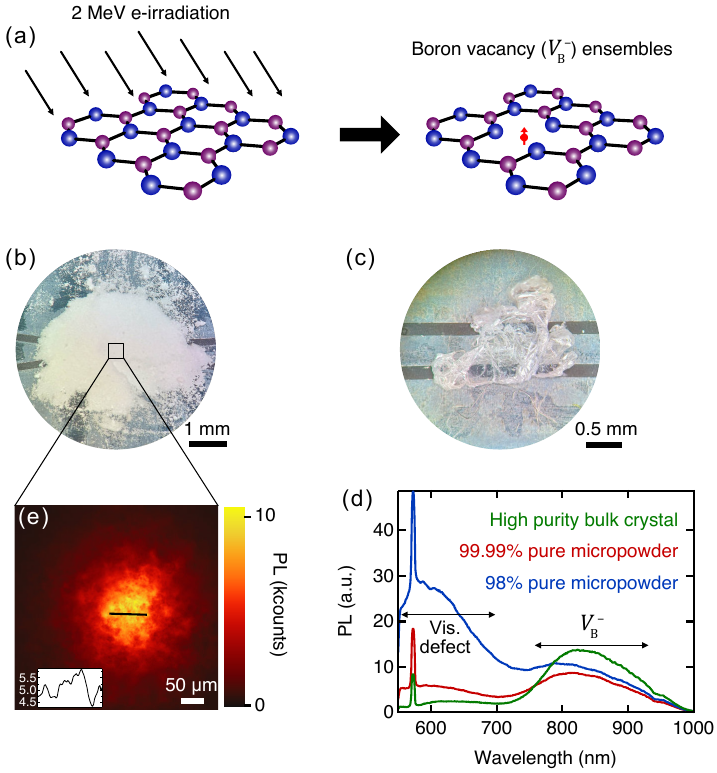}
\caption{(a) Schematic of the electron irradiation process, resulting in the creation of $V_{\rm B}^-$ (red spin) defects in hBN. (b) hBN micropowder deposited on a MW stripline for measurement. (c) Bulk hBN crystal on a MW stripline for measurement. (d) PL spectra taken under CW 532 nm illumination for three samples of differing initial purity following an identical electron irradiation procedure (2~MeV, $5\times 10^{18}$~cm$^{-2}$). The approximate PL bands belonging to ``visible" defects and near-infrared emission from $V_{\rm B}^-$ defects are marked. A similar level of $V_{\rm B}^-$ fluorescence is evident in each case but the initial purity of the crystal is linked to the level of $<700$~nm fluorescence. (e) Widefield $V_{\rm B}^-$ PL image of an irradiated hBN micropowder. The PL received is roughly uniform across the centre of the laser spot, shown by the inset linecut.}
\label{fig:intro}
\end{figure*}

Here we focus on the lower defect density regime (vacancy densities up to around 100~ppm), produced by electron irradiation of doses ranging from $2\times 10^{16}$ to $2\times 10^{19}$~cm$^{-2}$, and attempt to observe an optimum with respect to two prominent measurement modalities, ODMR and $T_1$ relaxometry. We find the defect density generated by a dose of $5\times 10^{18}$~cm$^{-2}$ is optimal for a high purity micropowder, but that the trends appear different for other samples, suggesting a dependence on the existing crystal purity. These results suggest that the lower defect densities produced by this level of electron irradiation may produce more sensitive spin ensembles than the high-dose ion implantation most commonly considered in the literature so far (which can approach anticipated levels of vacancy creation nearing 1\%~\cite{Healey2022a}).

\section{Methods}
\subsection{Sample preparation}

A range of hBN samples were subjected to 2~MeV electron irradiation at various doses, as depicted schematically in Fig.~\ref{fig:intro}(a). As in previous, lower-dose studies~\cite{NgocMyDuong2018}, the irradiation was performed under ambient conditions using a Cockcroft-Walton 2 MV 60 kW electron accelerator. The hBN samples were wrapped in aluminium foil and mounted on a water cooled copper plate. 

Micropowders of two purities which we refer to as high (3-4~$\mu$m particle size, 99.99\% purity from SkySpring Nanomaterials) and low purity (5~$\mu$m particle size, 98\% purity from Graphene Supermarket) respectively [see optical image Fig.~\ref{fig:intro}(b)], were selected to probe the effect of other contamination as well as irradiation on spin properties. Large particle sizes, effectively reflecting the bulk regime, were chosen to isolate these factors from surface effects~\cite{Durand2023}. We also irradiated high purity, bulk crystals [Grade A from HQ Graphene, pictured in Fig.~\ref{fig:intro}(c)] and exfoliated individual flakes (sub-$\mu$m thickness) to demonstrate that the technique is applicable to applications that require custom sensors with large lateral size. 

\subsection{PL collection and analysis}
In principle, these samples differ only in their purity, and by subjecting them to identical electron irradiation procedures we may expect a similar density of $V_{\rm B}^-$ defects to be produced. To assess the levels of $V_{\rm B}^-$ creation, we collected photoluminescence (PL) spectra using an Ocean Insight Maya2000 Pro spectrometer, using typical illumination conditions for widefield $V_{\rm B}^-$ sensing (532~nm, peak power density $\approx 4.5$~kW/cm$^2$). In Fig.~\ref{fig:intro}(d) we plot PL spectra obtained for representative samples from each category, irradiated with a dose of $5\times 10^{18}$~cm$^{-2}$. We can see that the near-infrared fluorescence associated with the $V_{\rm B}^-$ defect (with a characteristic phonon side band (PSB) peaking at $\approx 800$~nm) is indeed of similar amplitude in each case, but that PL exhibited below 700~nm (which we refer to as ``visible" emission henceforth) is highly variable. This visible fluorescence is attributed to a class of emitters~\cite{Stern2022,Guo2021,Scholten2023} which, although their precise nature is unknown, are believed to be related to carbon impurities~\cite{Mendelson2021}. We leave investigations into optimised creation of these defects for future work, although we note that visible PL was observed to increase with irradiation dose~\cite{NgocMyDuong2018}.

Although we do not consider the spin properties of these visible defects in this work, it can be seen that the tail of the visible PSB overlaps with the $V_{\rm B}^-$ emission and hence will reduce readout contrast. By filtering our PL collection (750~nm long pass filter), we are able to mostly isolate the $V_{\rm B}^-$ signal, however we note that the variable tail of visible defect emission will add some variability to, for example, the measured PL contrast in optically detected magnetic resonance (ODMR) measurements. For our PL measurements of the micropowders we use films produced by dispersing the powder in isopropanol, dropcasting, and allowing the isopropanol to evaporate~\cite{Scholten2023}. This procedure results in a relatively uniform film (see PL image in Fig.~\ref{fig:intro}(e), where the Gaussian shape is from the laser spot) that can be made much thicker than the expected extent of our laser spot, and hence all samples are effectively infinite in extent. 

For determining shot-noise-limited sensitivity, it is more appropriate to use the PL collection rate as would be captured in a practical experiment. For our setup (introduced below), this is the count rate as measured by an sCMOS camera with a collection window $>750$~nm. 
\subsection{Spin measurements}
\begin{figure*}[hbt]
\centering
\includegraphics[scale=1]{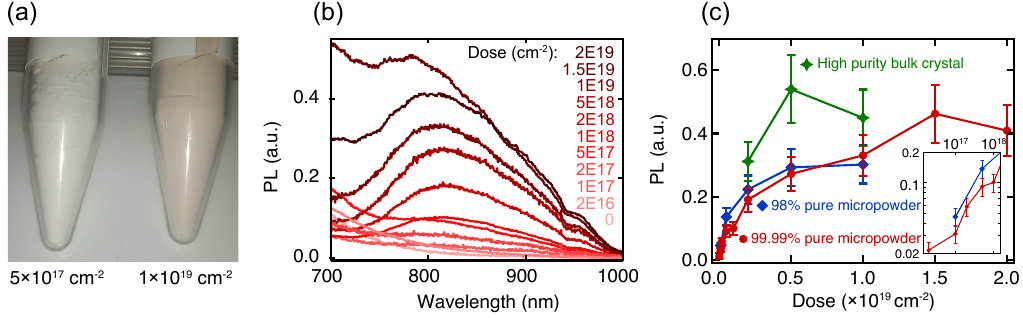}
\caption{(a) Photograph of two vials of hBN micropowder irradiated with different 2~MeV electron doses. The higher dose (right) has a noticeably red tint. (b) PL spectra recorded under CW 532~nm laser illumination for the high purity micropowder series. The characteristic $V_{\rm B}^-$ PSB peaking around 850~nm is visible for all irradiated samples, with a variable tail from visible defect emission also evident. (c) $V_{\rm B}^-$ PL extracted from PL spectra, taking values at a constant wavelength of 830~nm. Data is presented for a lower purity (98\%) micropowder and a high purity bulk crystal (HQ Graphene) as well as the samples shown in (b). The error bars are set to 20\% of the values to reflect observed variation across ensembles of the order of the maximum pixel-to-pixel variation seen in the linecut in Fig.~\ref{fig:intro}(e), which was similar to the variation in PL spectrum intensity measured at different spots. Inset is a log/log plot of the PL for doses of $1\times 10^{18}$~cm$^{-2}$ and less.}
\label{FigPL}
\end{figure*}
Spin measurements were taken on a purpose-built room temperature widefield microscopy setup. A 532~nm laser (Quantum Opus) provides spin initialisation and readout, gated by an acousto-optic modulator (Gooch \& Housego R35085-5). PL is collected through a microscope air objective (Nikon 20x S Plan Fluor ELWD, NA=0.45) and onto an sCMOS camera (Andor Zyla). RF radiation to manipulate the spin states is supplied by a Windfreak SynthNV PRO signal generator and amplified (Mini-Circuits HPA-50W-63+). Camera exposures, laser pulses, and RF pulses are all sequenced using a SpinCore Pulseblaster ESR 500 MHz card. 

For spin measurements, hBN powders were deposited on a silver stripline (0.4~mm) dry. Pulsed ODMR sequences utilised a constant $\pi$ time of 40~ns (12.5 MHz) and 500~ns laser pulses. CW ODMR sequences used a reduced Rabi frequency of 5 MHz. $T_1$ sequences used laser pulse durations of 1~$\mu$s. 

For all spin measurements we utilised an input laser power of $\approx 350$~mW,  focussed to a spot size of $\approx 70$~$\mu$m diameter, for a peak laser power density of $\approx 4.5$~kW/cm$^2$. 

\section{Results}
\subsection{$V_{\rm B}^-$ photoluminescence} 
We begin by assessing the scaling of $V_{\rm B}^-$ PL signal with electron irradiation dose. The as-received samples do not exhibit any $V_{\rm B}^-$ fluorescence, and become fluorescent above 750~nm following irradiation. For the powdered samples, the high levels of defect creation  following high doses are accompanied by an increased red tint, as shown in Fig.~\ref{FigPL}(a). In Fig.~\ref{FigPL}(b) we show the PL spectra (focussing on the $V_{\rm B}^-$ emission band now) for one set of samples, the high purity micropowder, which we irradiated with the widest range of doses. It can be seen that the lowest dose ($2\times 10^{16}$~cm$^{-2}$) is successful in activating a $V_{\rm B}^-$ ensemble, with the PL increasing with higher doses. 

We summarise the PL signal observed for all samples in Fig.~\ref{FigPL}(c), where we plot the peak $V_{\rm B}^-$ PL at $\sim 830$~nm for simplicity. At this wavelength, the influence of the visible PL tail is largely suppressed and so the trends plotted are attributed solely to the $V_{\rm B}^-$ defect, however we note that the visible PL tail is more significant for the lower purity micropowder.

It can be seen from Fig.~\ref{FigPL}(c) that the initial, roughly linear, increase in $V_{\rm B}^-$ PL with irradiation dose slows past roughly $2\times 10^{18}$~cm$^{-2}$, indicating we approach saturation. Assuming we predominantly create single vacancies (which is expected for these irradiation doses), the observed sub-linear scaling at higher doses may indicate that the proportion of boron vacancies that are negatively charged decreases with increasing dose, in line with recent observations made in the high-defect density regime~\cite{Gong2022}. The high purity bulk crystal exhibits higher PL than the powders which may be due to the higher quality material offering a more favourable Fermi level position for $V_{\rm B}^-$ formation. However, we also cannot discount an optical contribution; the powder deposition [see Fig.~\ref{fig:intro}(b)] likely results in more scattering which could impact laser delivery and PL collection. 

We do not have a direct measurement of vacancy creation in our samples, however, Murzakhanov \textit{et al.}~\cite{Murzakhanov2021} estimated a $V_{\rm B}^-$ density of 5.7~ppm via an electron paramagnetic resonance (EPR) measurement, created by 2~MeV electron irradiation at a dose of $6\times 10^{18}$~cm$^{-2}$ in a bulk hBN crystal. This is in broad agreement with the expectation based on previous empirical and Monte Carlo studies into electron irradiation of materials such as diamond~\cite{Campbell2002} and recent quantification of relevant energy scales in hBN~\cite{Bui2023}, the combination of which suggests that a 2~MeV electron should produce order $1\times 10^{-4}$~vac./$\mu$m on average. This $V_{\rm B}^-$ density is comparable to the value reported following neutron irradiation in Gottscholl \textit{et al.}~\cite{Gottscholl2021a} (also measured by EPR), and is an order of magnitude lower than the densities estimated in Gong \textit{et al.}~\cite{Gong2022} for He$^+$ irradiation. In this context, the range of irradiation doses considered in our work is expected to range from among the lowest defect creation densities considered so far in the literature for ensemble $V_{\rm B}^-$ creation, well below 1~ppm, to those achievable by low-dose ion implantation (above 100~ppm vacancy production).

\subsection{ODMR properties and DC sensitivity} 
\begin{figure*}[htb]
\centering
\includegraphics[scale=1]{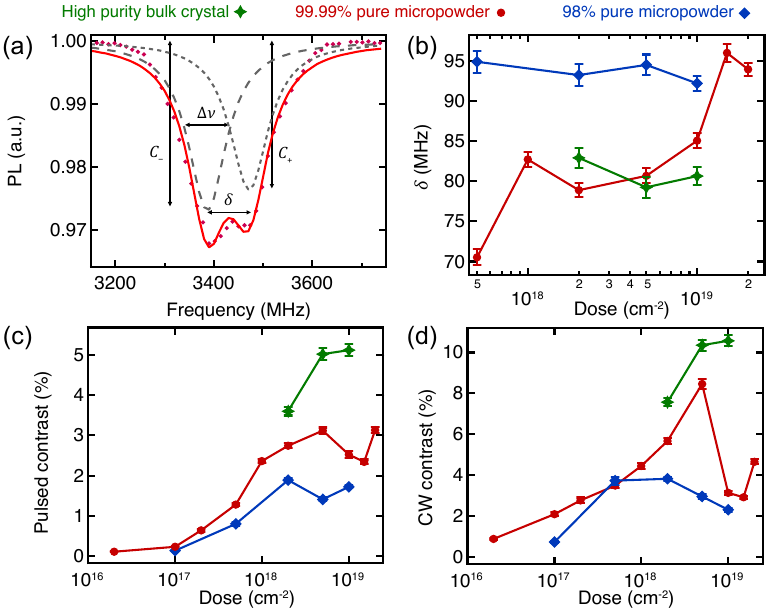}
\caption{(a) Pulsed ODMR spectrum taken at zero magnetic field (high-purity micropowder, irradiation dose $1\times 10^{19}$~cm$^{-2}$). The dots are the experimental data and the solid red line is a fit based on the sum of two Lorentzian lineshapes (grey dashed traces). These Lorentizans are centred at frequencies $f_-$ and $f_+$, have amplitudes $C_-$ and $C_+$ and a shared full width at half maximum (FHWM) $\Delta \nu$. (b) The splitting between $f_+$ and $f_-$, $\delta$, which is related to the electric field $E$. $\delta$ is related to the charged defect density and is accordingly seen to increase with irradiation dose. A correlation with the initial crystal purity is also evident. The extraction of ODMR splitting for samples irradiated with less than $5\times 10^{17}$ was unreliable and so we do not include these measurements. (c) Pulsed ODMR contrast versus irradiation dose. The trends between samples are very different, indicating a strong dependence on initial crystal purity. The contrast plotted here is the average of the two fit amplitudes $C_-$ and $C_+$. (d) CW ODMR constrast. All error bars represent the standard errors from the curve fitting.}
\label{Fig3}
\end{figure*}
Ultimately, we are primarily concerned with the sensitivity of a spin-based measurement, which absolute PL will be but one factor in. Other spin properties will play into the measurement sensitivity, which we characterise on a typical widefield sensing apparatus. Two measurement modalities have mainly been used in $V_{\rm B}^-$-based sensing so far, ODMR and $T_1$ relaxometry, which we will consider in turn. 

In Fig.~\ref{Fig3}(a) we show a typical pulsed ODMR spectrum under zero magnetic field, for a high purity micropowder sample following a dose of $1\times 10^{19}$~cm$^{-2}$. ODMR spectra such as this can be fit using two Lorentzian lineshapes with a shared width $\Delta \nu$ and individual amplitudes (contrasts $\mathcal{C}_-$ and $\mathcal{C}_+$) and central frequencies ($f_-$ and $f_+$), where the prescription $\pm$ refers to the resonance belonging to the transition between the two spin transitions within the ground state. The resonance contrasts and linewidths are relevant in determining measurement sensitivity (as we will see below) and the resonance locations can be linked to the quantities ${D = (f_+ + f_-)/2}$, the zero-field splitting parameter, and ${\delta =(f_+ - f_-)}$, which may be expected to scale with irradiation dose as they can be related to lattice strain and charged defect density respectively~\cite{Lyu2022,Gong2022}. As well as pulsed ODMR, wherein the excitation/readout laser is temporally separated from the MW drive, it is also possible to conduct continuous wave (CW) ODMR, where laser and MW are applied simultaneously. Despite incurring additional power broadening~\cite{Dreau2011}, CW ODMR often results in higher sensitivity measurements for $V_{\rm B}^-$ sensing, however can also produce sharp increases in temperature that can be undesirable~\cite{Healey2022a}. In the following, we will consider both pulsed and CW protocols for extracting sensitivities and restrict ourselves to pulsed ODMR for the extraction of the zero-field resonant frequencies. 

We start by looking at the $\delta$ parameter in Fig.~\ref{Fig3}(b). There is a clear increasing trend evident for the high-purity micropowder series, for which doses over nearly two orders of magnitude are considered, while the trend is less stark in the other two series. This quantity has previously been related to a charged defect concentration in the higher density regime with the proportionality $\delta \propto 2 d_{\perp} E_{\perp}$~\cite{Gong2022,Udvarhelyi2023}, where $E_{\perp}$ is the transverse electric field and $d_{\perp}\approx 35$~Hz/V.cm$^{-1}$ is the $V_{\rm B}^-$ transverse electric field susceptibility~\cite{Gong2022}. Comparing to Gong \textit{et al.}~\cite{Gong2022}, the values of $\delta$ corresponding to the highest doses studied are consistent with charged defect densities approaching 100~ppm, as expected. In the ideal case, the number of charged defects present in the crystal would be twice the $V_{\rm B}^-$ density (half belonging to the $V_{\rm B}^-$ defects and half to the counterpart positively-charged electron donors), however it is also possible that the irradiation produces other charged defects, for instance charged vacancies on nitrogen sites and more complex defects. 
The larger $\delta$ values measured for the lower-purity micropowder corroborate this expectation, likely being due to a higher pre-existing charged defect density, including that related to the strong visible emission observed in Fig.~\ref{fig:intro}(d). Previous measurements have estimated the paramagnetic defect density of similar hBN powders at above 10~ppm~\cite{Scholten2023}, which matches the current data. 

Moving to ODMR contrast, we see some evidence for differing trends versus irradiation dose based on the purity of the starting material. Starting with a high power pulsed (40~ns $\pi$ pulse) protocol in Fig.~\ref{Fig3}(c), we see that, similar to the PL data, the high-purity micropowder series features a peak in contrast at $5 \times 10^{18}$~cm$^{-2}$, while the lower purity powder's contrast peaks at $2 \times 10^{18}$~cm$^{-2}$ and the bulk crystal exhibits a higher contrast that does not appear to fully saturate versus irradiation dose up to $1 \times 10^{19}$~cm$^{-2}$. The difference between the three series [which we note is separate to the PL trends in Fig.~\ref{FigPL}(c)] may be due to an increase in PL from other defects or a reduction in spin pumping arising from a charge migration effect in the higher defect density regime~\cite{Manson2018,Capelli2022}. 

In Fig.~\ref{Fig3}(d) we plot the CW ODMR contrast, which features similar trends with a crucial difference: a more sudden drop in contrast after $5 \times 10^{18}$~cm$^{-2}$ for the high-purity micropowder, accompanied by a steady drop versus dose in the low-purity powder. The cause of the exacerbation of the trends observed in the pulsed ODMR case is unclear, but may support the interpretation that complicated charge dynamics during laser illumination are relevant. We also note that the low-purity micropowder, with its more significant visible PL and electric field splitting, exhibits the lowest PL contrast and tends to decrease with dose from a maximum at $2 \times 10^{18}$~cm$^{-2}$, especially in the CW series, while the high purity bulk crystal contrast strictly improves with irradiation dose over the range considered. Although not definitive, these trends suggest that the additional defects in the lower purity materials, both native and activated by irradiation, contribute to a lower $V_{\rm B}^-$ spin contrast beyond simply adding background PL. 

\begin{figure*}
\centering
\includegraphics[scale=1]{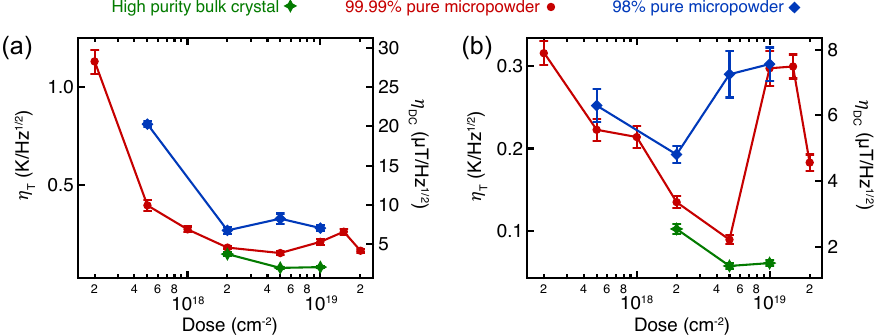}
\caption{(a) Temperature (left axis) and DC magnetic field (right) sensitivity calculated for the different samples, based on pulsed ODMR measurements. Here we consider PL collected from a $\approx 70 \times 70~\mu$m$^2$ region under 350~mW laser excitation, collecting all PL above 750~nm. (b) As (a) but using CW ODMR measurements. Note we exclude data from doses lower than $2\times 10^{17}$~cm$^{-2}$ for plot readability.}
\label{fig:sensitivity}
\end{figure*}

To benchmark the quality of the ensembles produced for ODMR-based sensing, we consider the sensitivity to temperature shifts at zero field, which is given by~\cite{Barry2020}

\begin{equation}
\eta_{\rm T} = \frac{4}{3\sqrt{3}}\frac{2 \pi}{\alpha}\frac{\Delta \nu}{\mathcal{C}\sqrt{\mathcal{R}}},
\label{tempsens}
\end{equation}

where $\alpha = 0.70$~MHz/K is the $V_{\rm B}^-$ temperature coefficient when linearising the response near room temperature~\cite{Healey2022a,Gottscholl2021a,Liu2021}, $\Delta \nu$ is the power broadened ODMR linewidth, $\mathcal{C}$ is the ODMR contrast, and $\mathcal{R}$ is the photon count rate, which is effectively equal to the value under CW laser illumination given the high duty cycle in even our pulsed measurements. Note that we focus on temperature sensitivity here since we mainly consider ensembles of powders that feature randomly-oriented defects on average and hence are not suited to measuring magnetic fields, however the DC magnetic field sensitivity can be obtained by multiplying Eqn.~\ref{tempsens} by $\alpha / \gamma_e$, where $\gamma_e$ is the electron gyromagnetic ratio. This quantity is displayed on the right axes of Fig.~\ref{fig:sensitivity}, however this is indicative only since away from zero field the hyperfine structure of the $V_{\rm B}^-$ defect will broaden its ODMR resonance more significantly, leading to a greater linewidth~\cite{Healey2022a}. 

The results are plotted in Fig.~\ref{fig:sensitivity}, where we separate the sensitivity extracted using pulsed [Fig.~\ref{fig:sensitivity}(a)] and CW [Fig.~\ref{fig:sensitivity}(b)] ODMR. The sensitivity values reported are based on PL collection from illuminated regions approximately 70~$\mu$m in diameter, using a 20$\times$, NA = 0.45 objective. The trends are similar in both cases (mostly following the trends in contrast as this quantity was the most variable), with optimal sensitivities obtained with a dose of $5\times 10^{18}$~cm$^{-2}$ in the case of the higher purity crystals, and the lower dose of $2\times 10^{18}$~cm$^{-2}$ for the lower purity powder. The sensitivities obtained using pulsed and CW protocols in this case are similar at the optimal doses, with the CW protocol performing better on average, although we note that the pulsed protocol can avoid laser heating of a sample and so will likely be preferred in practice. 

\subsection{$T_1$ and relaxometry sensitivity}
\begin{figure*}[htb]
\centering
\includegraphics[scale=1]{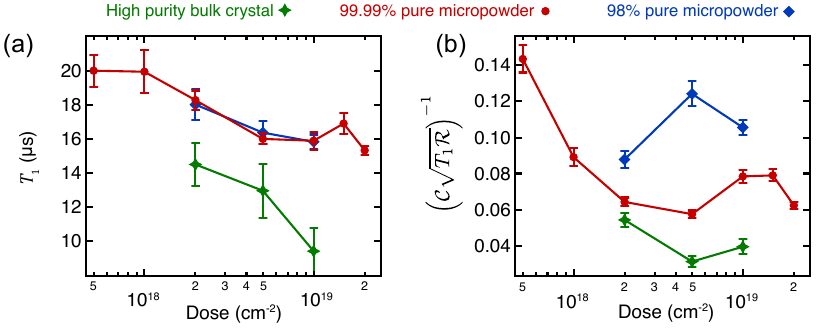}
\caption{(a) $T_1$ values recorded versus irradiation dose. $T_1$ decreases with irradiation dose, likely due to the introduction of magnetic noise and possibly due to modification of phonons. No dependence on the initial crystal purity is evident. (b) Figure of merit proportional to sensitivity (inverse to SNR) of a relaxometry measurement. The difference between the two powders is primarily due to the measurement contrast since $T_1$ and the PL count rate are similar between the two series. An optimal sensitivity is again found for a dose of $5 \times 10^{18}$~cm$^{-2}$ in the case of the high purity micropowder and bulk crystal.}
\label{fig:T1}
\end{figure*}

Finally, we measure the spin relaxation time $T_1$, which sets the practical upper limit for spin coherence and the sensitivity of relaxometry experiments~\cite{Tetienne2013,Robertson2023}. Here we use longer laser pulse durations, $\approx 1~\mu$s (up from 500~ns), to ensure high quality and even initialisation into the $\ket{0}$ ground state~\cite{Robertson2023}. The results are plotted in Fig.~\ref{fig:T1}(a), and we find a consistent dependence on irradiation dose across the various data sets. The reduction we observe for the micropowders, down to about 15~$\mu$s for a dose of $2\times 10^{19}$~cm$^{-3}$, is again consistent with previous ion implantation results~\cite{Gong2022,Guo2022}, again assuming that our level of defect creation is at the 100~ppm level and below. This result implies that additional damage introduced by the irradiation is a relatively minor source of accelerated spin relaxation compared to the processes that limit room temperature $T_1$ to $\approx 20$~$\mu$s, and there is no apparent dependence on crystal purity. Given the importance of high initialisation for accurate $T_1$ measurements, we measure individual flakes exfoliated from the bulk crystal, rather than the parent crystal as before. 

The mechanism behind the limiting room temperature $V_{\rm B}^-$ $T_1$ and its previously observed temperature scaling is still an open question, but could include phonon and thermally-activated magnetic noise components~\cite{Mondal2022}. Given our observation that the $T_1$ trends appear to be insensitive to the pre-existing charged defect density [c.f. Fig.~\ref{Fig3}(b)], one possible inference could be that the irradiation process modifies the phonon spectrum through the lattice damage introduced~\cite{Zhang2021b}. The $T_1$ values recorded for flakes exfoliated from bulk crystals are consistently lower than the micropowders, although they are typical of exfoliated flakes measured elsewhere in the literature~\cite{Durand2023}. One interpretation in our case could be that the smaller hBN particles heat up during irradiation and effectively anneal, so that the vacancy density produced in the bulk crystals is actually higher. This conclusion may be supported by the higher levels of PL and spin contrast observed in the bulk crystals compared to micropowders with the same irradiation dose, however this is only a speculative conclusion at this stage. 

Again assuming a shot-noise-limited measurement, the sensitivity of a $T_1$ relaxometry measurement will be proportional to~\cite{Robertson2023}

\begin{equation}
\eta_{T_1} \sim \left(\mathcal{C}\sqrt{T_1 \mathcal{R}}\right)^{-1},
\end{equation}

where $\mathcal{C}$ is the contrast of the $T_1$ measurement, proportional to the pulsed ODMR contrast. Note that this is a unitless quantity and that the sensitivity of a $T_1$ relaxometry measurement will also be dependent on the distance between sensor and target, as well as the properties of the magnetic noise being sensed~\cite{Tetienne2013}. 
This figure of merit is plotted in Fig.~\ref{fig:T1}(b), where we find similar scaling and hierarchy as in Fig.~\ref{fig:sensitivity}. An irradiation dose of $5 \times 10^{18}$~cm$^{-2}$ appears optimal, except for the low-purity micropowder due to the trends in spin contrast -- overall the trend in $T_1$ are too subtle to overcome the more significant variation in contrast between the samples. Note we use the same PL values as before for the `high purity bulk crystal' data so that the comparison between data series is fair.

\section{Conclusion}
In this work we have found that electron irradiation is effective in activating $V_{\rm B}^-$ ensembles over a wide range of doses. Since electron irradiation carries the major advantage of scalability over ion implantation (which is depth-confined) and accessibility compared to neutron irradiation, we expect that the technique will be particularly useful for creating large amounts of quantum-active hBN nano and micropowders, and bulk crystals to facilitate repeated exfoliation of single crystal flakes for incorporation into van der Waals heterostructures.

We find that a dose of $5\times 10^{18}$~cm$^{-2}$ appears optimal for high purity crystals on our setup, while lower doses may be preferred for lower-quality starting material. If the defect density created by this dose is optimal regardless of the creation method, our results could imply that greater sensitivities can be obtained by considering lower density ensembles than have been prioritised in the literature so far, such as through high-dose ion implantation.

The dependence of sensitivity on initial crystal purity is strongly linked to background fluorescence from visible emitters that are present in a higher quantity in the lower purity powders. The most significant difference comes from a degradation in readout contrast, which is likely due to a combination of background fluorescence overlapping our PL collection window and an increased abundance of nearby defect sites that promote non-radiative decay pathways during illumination. The limiting ODMR linewidths and spin relaxation times, are not strongly linked to the hBN purity, being governed primarily by the dense nuclear spin bath and mechanical properties of the crystal respectively. We do, however, corroborate previous studies that found a  dependence of these properties on irradiation-induced lattice damage, confirming that this damage is significant even at the low end of defect creation. Future work could consider whether other treatments such as annealing under moderate temperatures~\cite{Suzuki2023} can produce more sensitive $V_{\rm B}^-$ ensembles and whether the optimal irradiation dose is changed.

\section*{Acknowledgements}
The authors thank Brett Johnson for fruitful discussions. This work was supported by the Australian Research Council (ARC) through grants CE200100010, FT200100073, FT220100053, DE200100279, DP220100178, DP220102518, and DE230100192, and by the Office of Naval Research Global (N62909-22-1-2028). P.R. acknowledges support through an Australian Research Council DECRA Fellowship (grant no. DE200100279) and an RMIT University Vice-Chancellor’s Senior Research Fellowship. I.O.R. is supported by an Australian Government Research Training Program Scholarship. S.C.S. gratefully acknowledges the support of an Ernst and Grace Matthaei scholarship. Part of this study was supported by QST President's Strategic Grant ``QST International Research Initiative". 
\bibliographystyle{apsrev} 
\bibliography{hbnirrad}
\end{document}